\documentclass[fleqn]{2020SCGE}

\usepackage[T1]{fontenc}
\usepackage{footmisc}
\usepackage{subfigure}
\usepackage{diagbox}
\usepackage{amsfonts,amsmath,amstext,amssymb,mathrsfs,gensymb}
\usepackage{makecell,multirow}
\usepackage{ulem}

\usepackage[figure,figure*]{hypcap}%??
\usepackage{booktabs}%???????
\usepackage{paralist}%??????
\usepackage{natbib}%??????
\usepackage{graphicx}%????
\usepackage{appendix}

\begin{document}
\ensubject{subject}
\ArticleType{Article}
\Year{2023}
\Month{May}
\Vol{}
\No{}
\DOI{}
\ArtNo{000000}
\ReceiveDate{2023-3-12}
\AcceptDate{2023-5-20}

\title{\vspace{-2 cm}Discovery of two rotational modulation periods from a young hierarchical triple system}{Discovery of two rotational modulation periods from a young hierarchical triple system} 

\author[1]{Yu-Tao Chen}{}%
\author[2,3,1,4]{Hai-Jun Tian \thanks{Corresponding author: hjtian@lamost.org}}{}
\author[5]{Min Fang}{}
\author[6,7]{Xiao-Xiong Zuo}{}%
\author[1]{Sarah A. Bird}{}
\author[1]{Di Liu}{}%
\author[1]{\\Xin-Yu Zhu}{}
\author[1]{Peng Zhang{}}{zhangpeng@ctgu.edu.cn} 
\author[1]{Gao-Chao Liu}{}%
\author[1]{Sheng Cui}{}

\AuthorMark{Chen Y T}
\AuthorCitation{Chen Y T, Tian H J, Fang M, et al}

\address[1]{~Center for Astronomy and Space Sciences, China Three Gorges University, Yichang, 443002, China}
\address[2]{~School of Science, HangZhou Dianzi University, HangZhou, 310018, China}
\address[3]{~Space Information Research Institute, Hangzhou Dianzi University, HangZhou, 310018, China}
\address[4]{~School of Physics and Astronomy, China West Normal University, NanChong, 637002, China}
\address[5]{~Purple Mountain Observatory, Chinese Academy of Sciences, Nanjing, 210023, China}
\address[6]{~National Astronomical Observatories, Chinese Academy of Sciences, Beijing, 100101, China}
\address[7]{~University of Chinese Academy of Sciences, Beijing, 100049, China.}

\abstract{GW~Ori is a young hierarchical triple system located in $\lambda$ Orionis, consisting of a binary (GW~Ori\,A and B), a tertiary star (GW~Ori\,C) and  a rare circumtriple disk. Due to the limited data with poor accuracy, several short-period signals were detected in this system, but the values from different studies are not fully consistent. As one of the most successful transiting surveys, the Transiting Exoplanet Survey Satellite (TESS) provides an unprecedented opportunity to make a comprehensive periodic analysis of GW~Ori. In this work we discover two significant modulation signals by analyzing the light curves of GW~Ori's four observations from TESS, i.e., 3.02 $\pm$ 0.15\,d and 1.92 $\pm$ 0.06\,d, which are very likely to be the rotational periods caused by starspot modulation on the primary and secondary components, respectively. We calculate the inclinations of GW~Ori\,A and B according to the two rotational periods. The results suggest that the rotational plane of GW~Ori\,A and B and the orbital plane of the binary are almost coplanar. We also discuss the aperiodic features in the light curves; these may be related to unstable accretion. The light curves of GW~Ori also include a third (possible) modulation signal with a period of 2.51$\pm$0.09\,d, but the third is neither quite stable nor statistically significant.}


\maketitle

\begin{multicols}{2}

\section{Introduction}
\label{sec:introduction}

GW~Ori is a young \citep[1.0 $\pm$ 0.1\,Myr,][]{2004AJ....128.1294C} pre-main sequence hierarchical triple system, located in $\lambda$ Orionis \citep[408 $\pm$ 10\,pc,][]{2021A&A...649A...6G}. The spectral type is G8 and it belongs to the classical T Tauri star-type \citep{2014A&A...570A.118F,2018ApJ...852...38P}. The system is composed of a binary (GW~Ori\,A and B, the former is the primary star) and a tertiary component (GW~Ori\,C). Initially, GW~Ori was found to be a spectroscopic binary with an inner orbital period of $\sim$242\,d (A-B) and a separation of $\sim$1\,AU \citep{1991AJ....101.2184M, 2017ApJ...851..132C}. Subsequently, the existence of GW~Ori\,C was confirmed by \cite{2011A&A...529L...1B} through near-infrared interferometry; the tertiary has an outer orbital period of $\sim$11\,years and a projected distance of $8-9$\,AU \citep[AB-C,][]{2017ApJ...851..132C}. \cite{2017ApJ...851..132C} measured the masses of GW~Ori\,A, B and C as 2.7\,M$_{\odot}$, 1.7\,M$_{\odot}$ and 0.9\,M$_{\odot}$, respectively. This result shows that the mass of GW~Ori\,C is significantly smaller than the other components. However, more recently \cite{2020Sci...369.1233K} found that GW~Ori\,B and C have almost the same mass (e.g., 2.47$\pm$0.33\,M$_{\odot}$, 1.43$\pm$0.18\,M$_{\odot}$ and 1.36$\pm$0.28\,M$_{\odot}$). GW~Ori has a huge circumtriple disk, where the dust is extended to $\sim$400\,AU and the gas is extended to $\sim$1300\,AU \citep{2017A&A...603A.132F,2020ApJ...895L..18B,2020Sci...369.1233K}. 
Atacama Large Millimeter Array (ALMA) observations indicate that the circumtriple disk is misaligned and warped \citep{2020ApJ...895L..18B,2020Sci...369.1233K}.
The orbits (A-B) and (AB-C) are inclined by 156$\pm$1\degree{} and 149.6$\pm$0.7\degree{} relative to the plane perpendicular to the line of sight \citep{2020Sci...369.1233K}, and misaligned by 44$\pm$5\degree{} and 54$\pm$7\degree{} with respect to the disk plane \citep{2017ApJ...851..132C}.

The period is one of the most important observable parameters for a binary or triple system, which opens a window to uncover the physical scenario of the system, e.g., the geometry, dynamical interaction, and so on. Besides the long-term orbital periods, some short-term (several days) periods have been reported in the literature. \cite{1990AJ.....99..946B} used {\it UBVRI} data from the European Southern Observatory (ESO) to obtain a 3.3\,d period through the ``string-length'' period-finding method \citep{1986A&A...165..110B,1989A&A...211...99B}. \cite{2014A&A...570A.118F} used the residual of the relative velocity from the Fiber-fed Extended Range Optical Spectrograph (FEROS) to obtain a period of $5-6.7$\,d by generalized Lomb Scargle \citep[GLS,][]{2009A&A...496..577Z}, that they interpreted as corresponding to the rotational period of GW~Ori\,A. Using a $V$-band light curve spanning 30 yr, \cite{2017ApJ...851..132C} revealed several new $\sim$30-day eclipse events $0.1-0.7$\,mag in depth and a 0.2\,mag sinusoidal oscillation that was clearly phased with the AB-C orbital period. After excluding eclipses, the authors recovered a period of 2.93$\pm$0.05\,d from 1.1 to 100 days in the high-cadence KELT data set \citep{2016MNRAS.459.4281K} utilizing the Lomb-Scargle (LS) algorithm \citep{1976Ap&SS..39..447L,1982ApJ...263..835S}; they claimed that this period corresponds to the rotational period of GW Ori\,A. Unfortunately, the values of these  short periods are not fully consistent with each other.

TESS provides a great opportunity to search for short-term periodicity and make a comprehensive periodic analysis for GW~Ori because its all-sky survey provides a nearly continuous series of full frame images (FFI) with a cadence of 30 minutes or 10 minutes (during its extended mission) for bright objects in the complete sky. Each 30-min (or 10-min) FFI provides precise ($\sim$5\,mmag) photometry within a combined field of view of 2300\,deg$^2$ (called a `sector') \footnote{https://heasarc.gsfc.nasa.gov/docs/tess/TESS-Intro.html}.

In this work, we use the TESS light curves to study the periodic modulations in the GW~Ori system and discuss the variability characteristics of the light curve. This paper is arranged as follows. In Section\,\ref{TESS}, we introduce the extraction and detrending of the GW~Ori light curves. In Section\,\ref{Period analysis}, we describe the periodic analysis performed on the TESS light curves. In Section\,\ref{Discussion}, we discuss the variability of the light curves. We summarize our results in Section\,\ref{Conclusions}.

%========================================================
%\section{GW~Ori in TESS}
\section{Extraction of TESS light curves}
\label{TESS}
GW~Ori has four observations in FFI (Figure\,\ref{GW_TPF}) useful for periodic analysis. The first observation was taken with a cadence of 30 minutes during the primary mission (Sector\,06), and the last three observations were taken with a cadence of 10 minutes during the extended mission (Sectors\,32,\,43,\,45). The specific dates \footnote{https://archive.stsci.edu/tess/tess\_drn.html} are shown in Table \ref{observationtime}.

\begin{table}[H]
\footnotesize
\begin{threeparttable}
\caption{TESS observation dates of GW~Ori}\label{observationtime}
\doublerulesep 0.1pt \tabcolsep 13pt
\begin{tabular}{ccc}
\hline
\multirow{2}{*}{Sector} & \multicolumn{2}{c}{BTJD (day)}          \\ \cline{2-3} 
                        & First           & Second          \\ \hline
06                      & 1468.27 - 1477.01 & 1478.11 - 1490.04 \\
32                      & 2174.22 - 2185.93 & 2186.94 - 2200.23 \\
43                      & 2474.16 - 2486.12 & 2487.18 - 2498.88 \\
45                      & 2525.50 - 2537.97 & 2539.14 - 2550.62 \\ \hline
\end{tabular}
\begin{tablenotes}
%\footnotesize
\item[]Notes: The first column is the sector number, and the second and third columns are the two observations. Data collection was paused for about 1.00 day between the orbits to download data. TESS Barycentric Julian Day (BTJD) is the format of recording time for TESS data products. This is a Julian day minus 2457000 Julian days and corrected to the arrival times at the barycenter of the Solar System, which is not affected by leap seconds \citep{2021AcASn..62...39T}.
\end{tablenotes}
\end{threeparttable}
\end{table}

Figure\,\ref{GW_TPF} shows the FFI of the four sectors of GW~Ori, which are downloaded through {\tt TESScut} \citep{2019ascl.soft05007B}. The image for each sector  is 10\,pixel $\times$ 10\,pixel. The target light curves are extracted from the masks (red shaded regions).

%===================================================

\subsection{Original light curves}
\label{Original light curves}
\begin{figure*}[htp]
  \centering
  \subfigure{\includegraphics[width=0.44\textwidth]{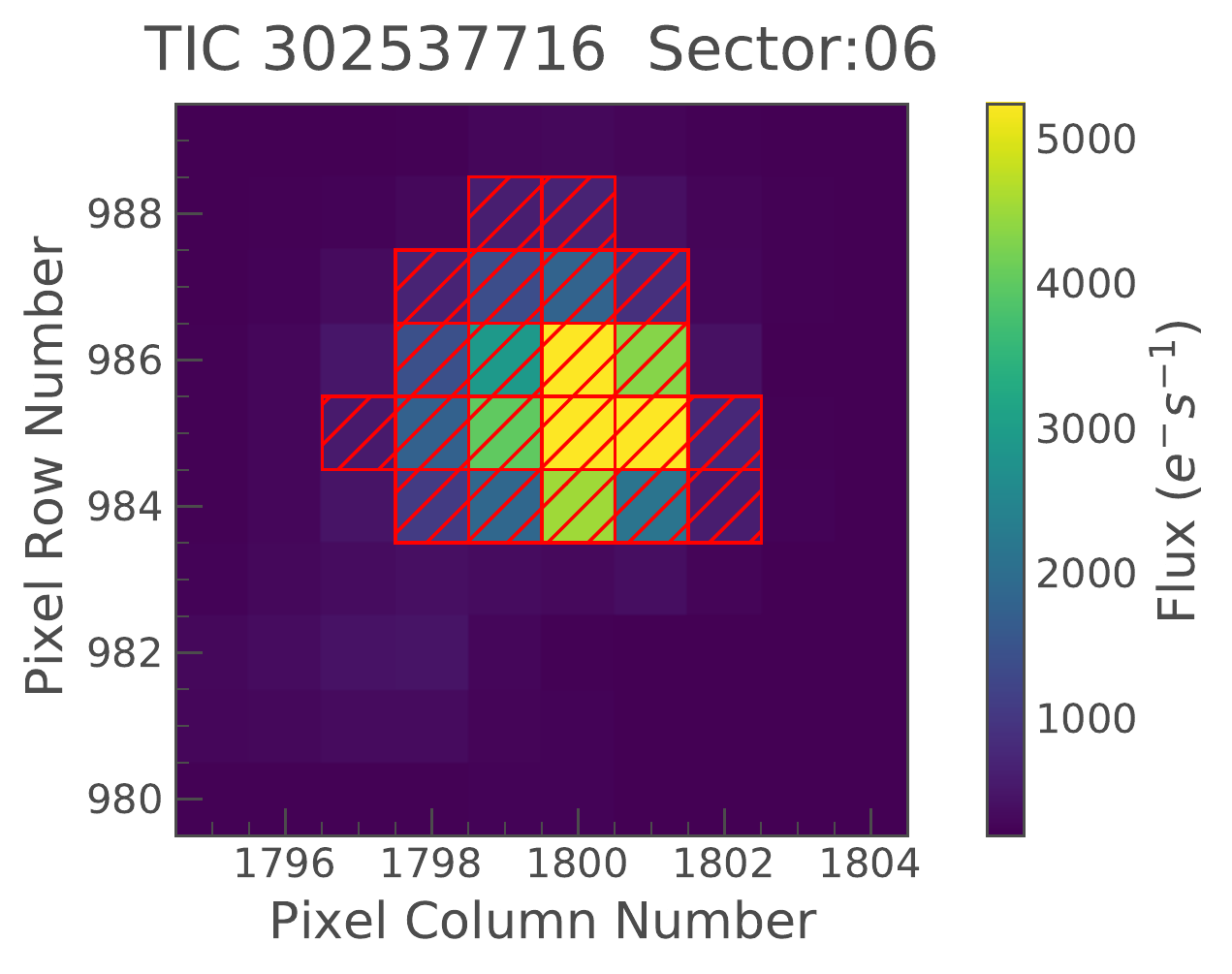}}
  \subfigure{\includegraphics[width=0.45\textwidth]{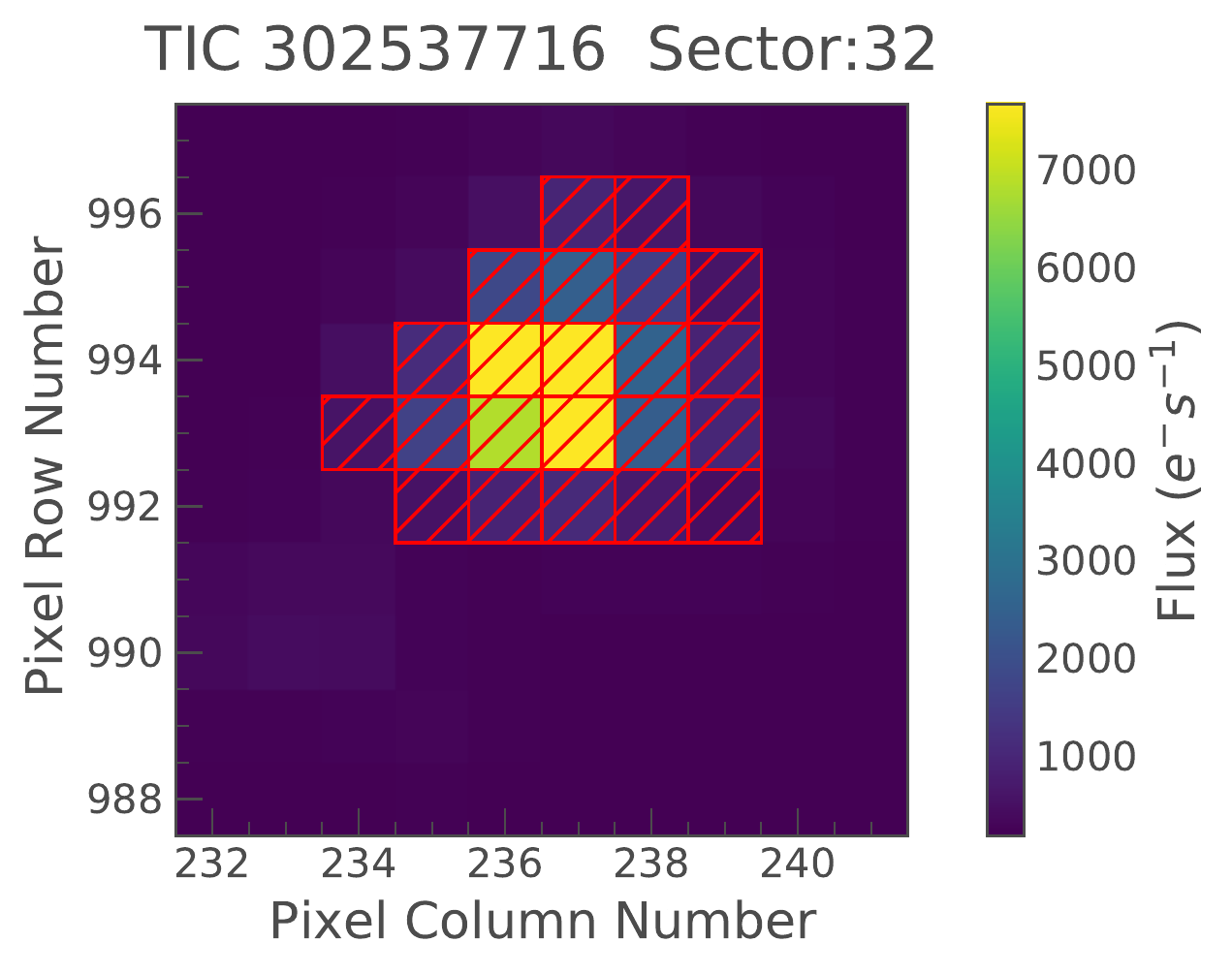}}
  \subfigure{\includegraphics[width=0.45\textwidth]{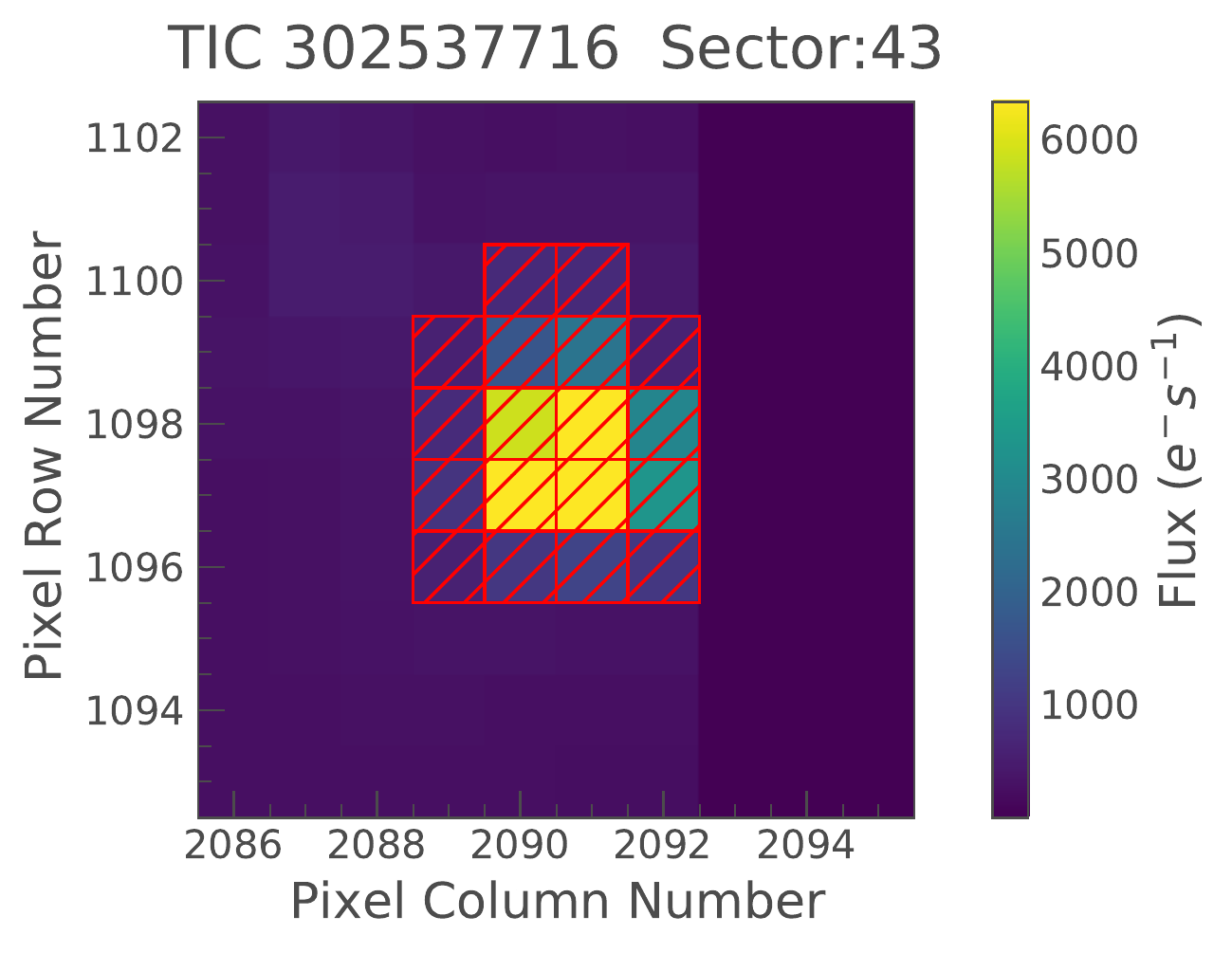}}
  \subfigure{\includegraphics[width=0.45\textwidth]{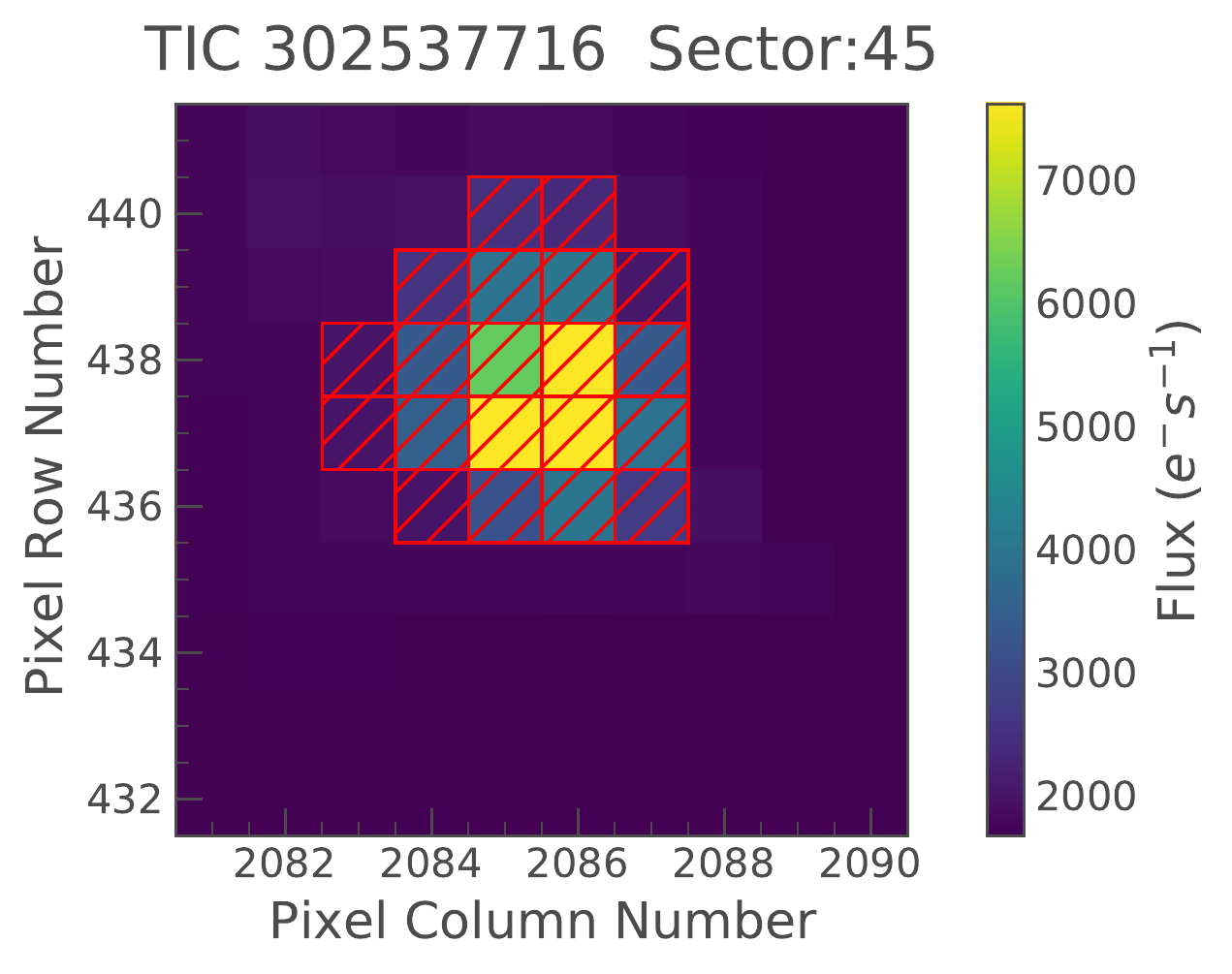}}
  \caption{TESS target pixel images of four sectors (06, 32, 43, and 45). Each image has a size of 10\,pixel $\times$ 10\,pixel. The pixels are color coded with flux (in electron-per-second) as represented in the color bar. The images of Sectors\,43 and 45 are rotated clockwise by 90\degree{} due to a camera direction adjustment. The red shaded regions are the optimal target masks made for the four sectors. The threshold values of the masks from Sector\,06 to Sector\,45 are 3, 3, 0.8, and 2, respectively.}%????
  \label{GW_TPF}%??????
\end{figure*}

We obtain the original TESS light curves using {\tt Lightkurve} \citep{2018ascl.soft12013L}. This is an open source python package, which provides a convenient method for users to access Kepler \citep{doi:10.1126/science.1185402} and TESS data products. We extract the original light curves of all four sectors using the area marked by the target masks (red shaded regions in Figure\,\ref{GW_TPF}) in FFI for further processing.

When downloading, we select data with the quality flag of 0 to ensure data reliability. The unit of the flux in a light curve is electron-per-second. We introduce our mask selections in FFI in Section \ref{mask} and our detrending method in Section \ref{detrending}.

%=====================================================
\subsection{Target Masks}
\label{mask}

Due to the 21 arcsec pixel resolution of TESS\footnote{https://heasarc.gsfc.nasa.gov/docs/tess/observing-technical.html}, the light curves have the possibility of being contaminated by other targets. An optimized mask ensures that the target has a high signal to noise ratio, and minimal contamination from the sources nearby in the projected sky. In this study, we adjust the masks using the function {\tt create\_threshold\_mask} provided in the package {\tt Lightkurve} \citep{2018ascl.soft12013L}.

\begin{figure}[H]
  \centering
  \includegraphics[width=0.5\textwidth]{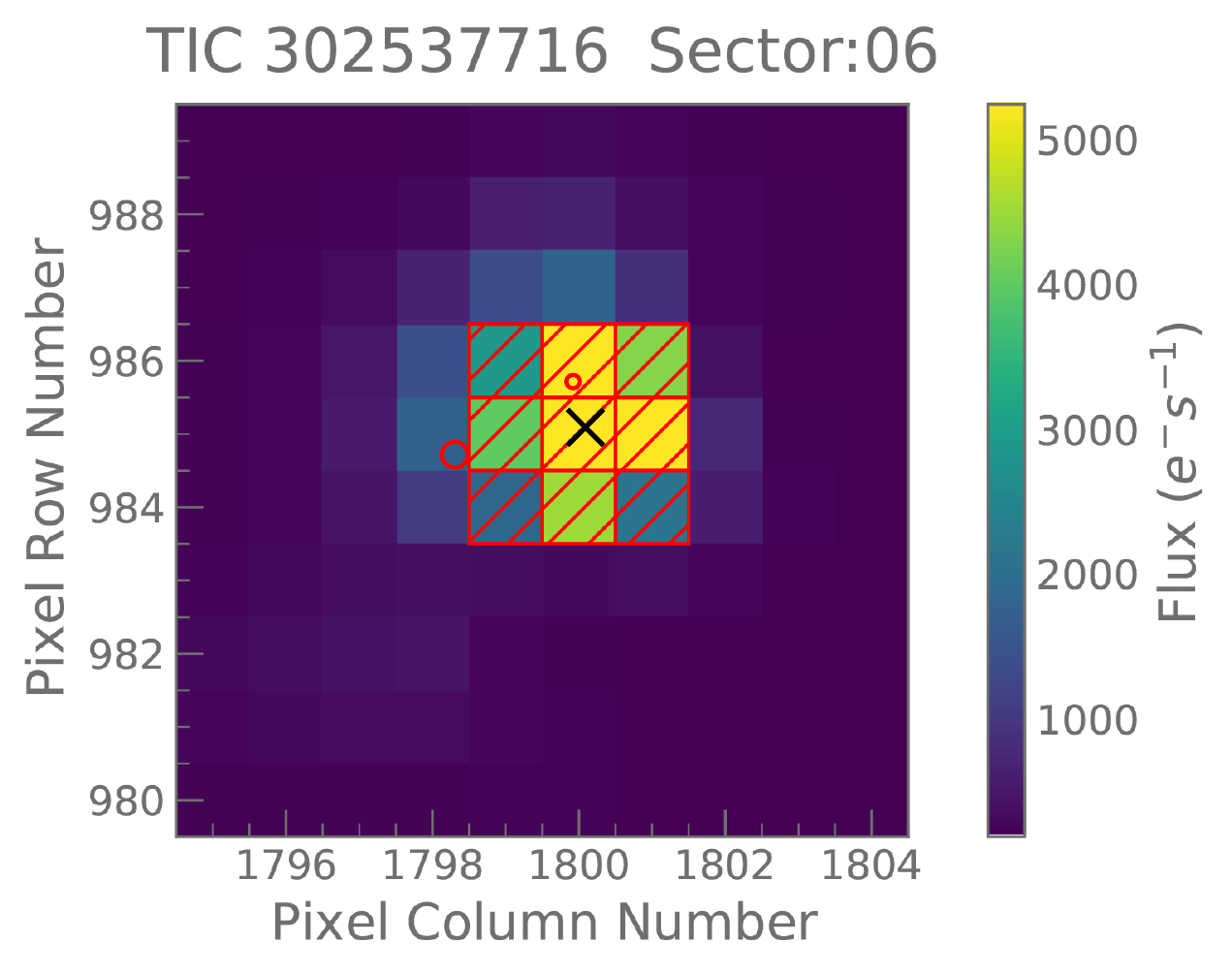}
  \caption{Pixel image of Sector\,06 under a red shaded mask with a threshold of 20, which avoids the brighter contaminator (12.7\,mag in $G$ band, marked with the larger red circle), but covers the dimmer contaminator (17.8\,mag in $G$ band, marked with the smaller red circle). Both of the contaminators are close in projected distance to our target GW~Ori (9.43 mag in $G$ band, marked with the black cross).}%????
  \label{lc_select}
\end{figure}

 In order to check whether there are bright stars near GW~Ori, we apply the function {\tt interact\_sky} to search for nearby stars in the image from {\it Gaia} DR2. We find two nearby {\it Gaia} sources, with IDs of 33408564989927229568 (12.7\,mag in $G$ band, marked by the larger red circle) and 334085646456658535392 (17.8\,mag in $G$ band, marked by the smaller red circle), near our target (GW~Ori, 9.43\,mag in $G$ band, marked by a black cross), as shown in Figure\,\ref{lc_select}. 

 We perform a check as to whether the contaminators have significant influence on the target light curve. We select a small mask for each of the four sectors (Sectors\,06, 32, 43, and 45 with threshholds of 20, 18, 8, and 23, respectively), in which the brighter contaminator is not covered (taking Figure\,\ref{lc_select} of Sector\,06 as an example). We then create the detrended light curves of GW~Ori (the detrending process of the light curve is described in Section \ref{detrending}), and compare them with those from a larger aperture with a threshold of 3, 3, 0.8, and 2 for each of the four sectors, respectively, in which the brighter contaminator is included. We find that the light curves with a larger mask in the four sectors have no significant differences compared to those with a smaller mask and only detect a slight difference in the first half of the light curve in Sector 06. These minute differences indicate that the contaminator does not have a significant impact on the light curve of GW~Ori The flux from the fainter star is far beyond the limiting magnitude ($9-15$\,mag with 50 ppm\footnote{Unit ppm is parts per million.} photometric precision) of TESS. In practice, the impact from the fainter star can be ignored.

After multiple tests, we finally select 3, 3, 0.8, and 2 as the optimal thresholds to make the masks for Sectors\,06, 32, 43, and 45, respectively, as shown by the red shaded areas in Figure\,\ref{GW_TPF}, and then proceed to create the light curves for all the sectors. 

%========================================================
\subsection{Light Curve Detrending}
\label{detrending}
 
The light curves extracted from the FFI have systematic trends caused by noise sources such as scattered light and instrumental effects; such trends are removed with the procedure of detrending. The package {\tt Lightkurve} provides a function to directly download the light curves which have been detrended through basis spline fits in the {\tt Quick Look Pipeline} \citep[QLP,][]{2020RNAAS...4..204H}. However, the light curves are detrended by batch processing more than 5 million stars in the QLP. For an individual source, the batch processing is probably not the best option. Considering that the light curve detrending is a crucial step for periodic analysis, we separately test two methods to detrend the light curves of GW~Ori.

\begin{figure*}[htp]
  \centering
  \subfigure[Sector\,06]{\includegraphics[width=8.9cm, trim=0.2cm 0.1cm 0.3cm 0.1cm, clip]{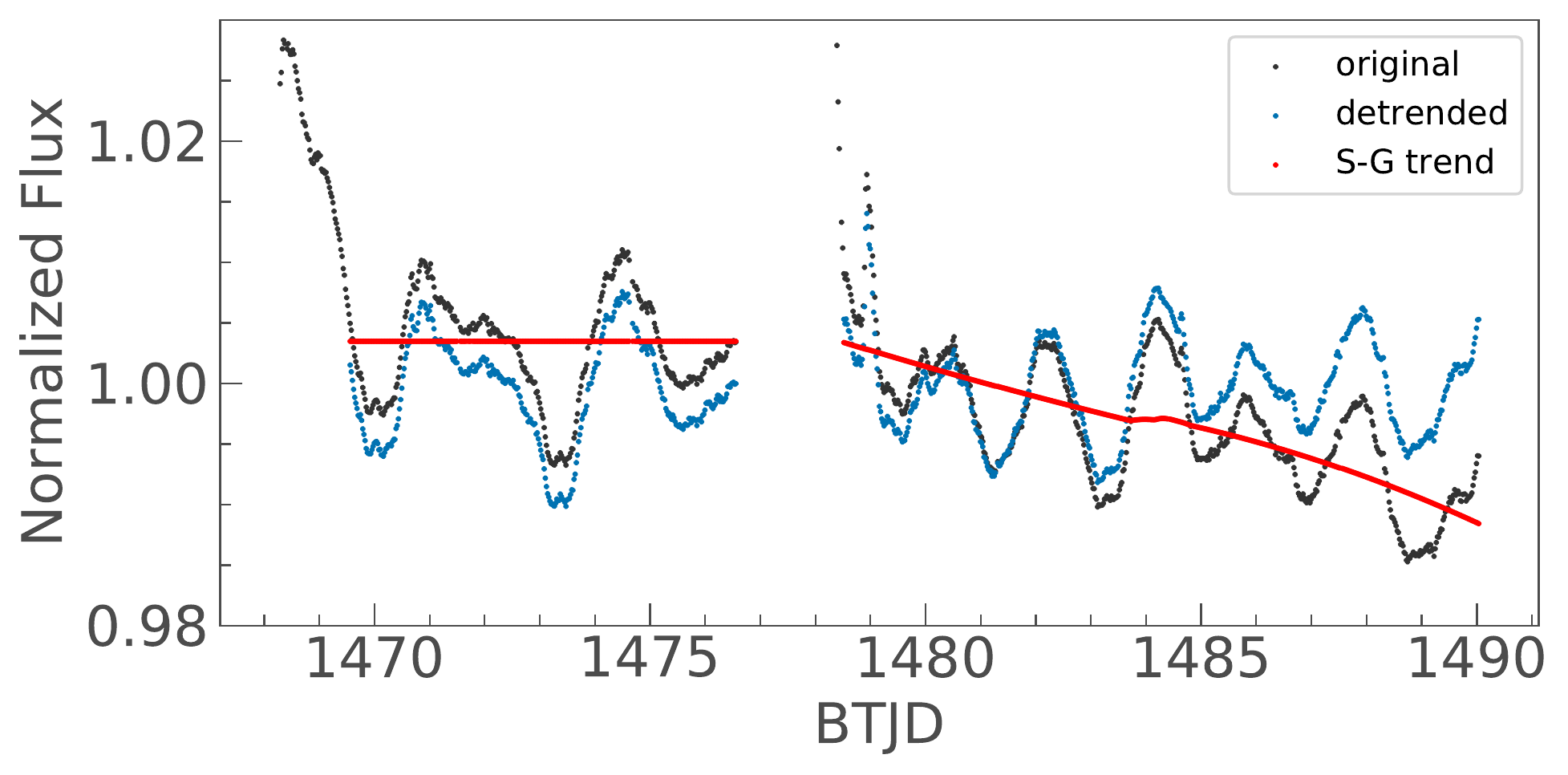}}
  \subfigure[Sector\,32]{\includegraphics[width=7.75cm, trim=2.7cm 0.1cm 0.2cm 0.1cm, clip]{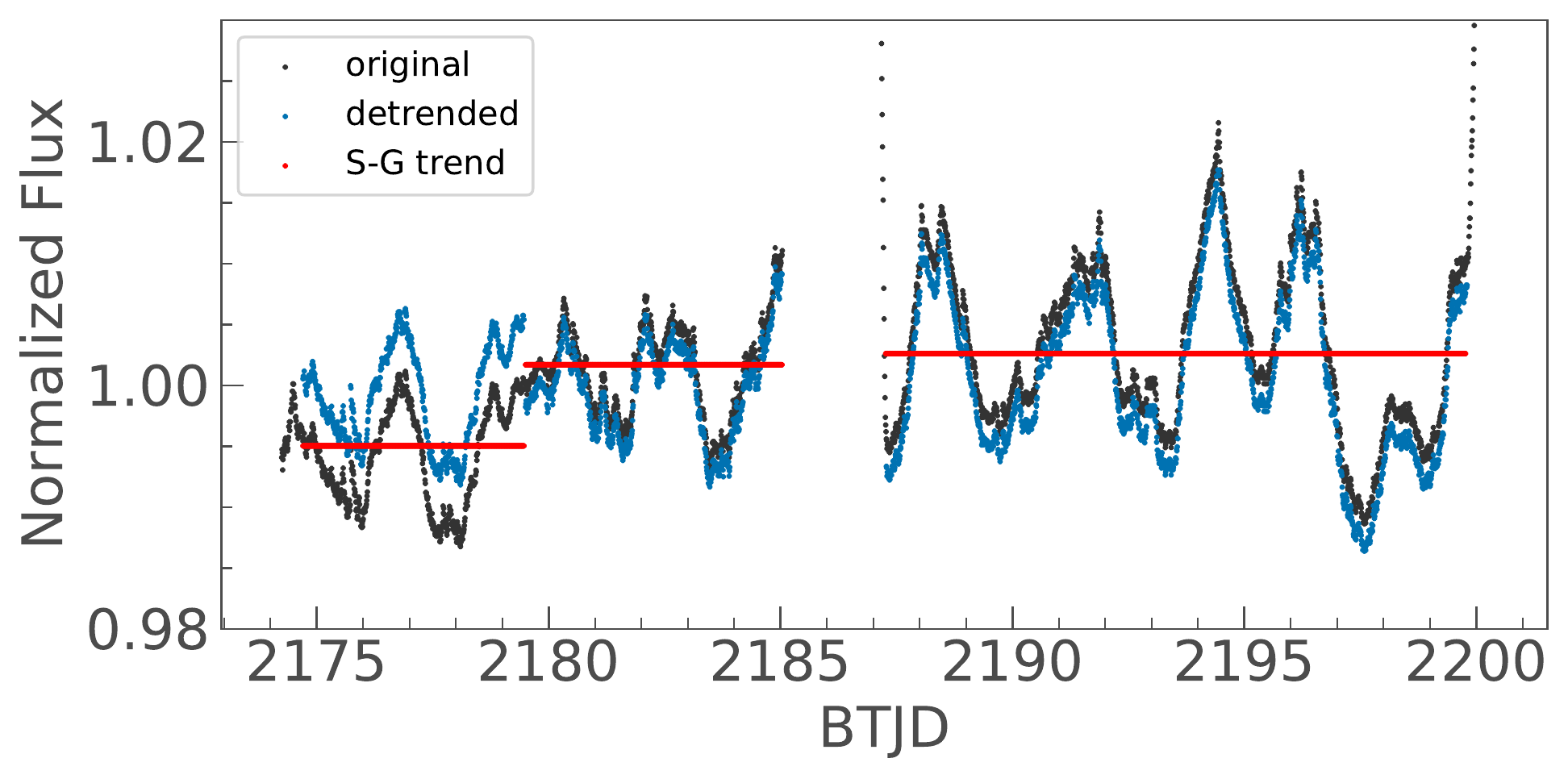}}
  \subfigure[Sector\,43]{\includegraphics[width=9.0cm, trim=0.2cm 0.1cm 0.2cm 0.1cm, clip]{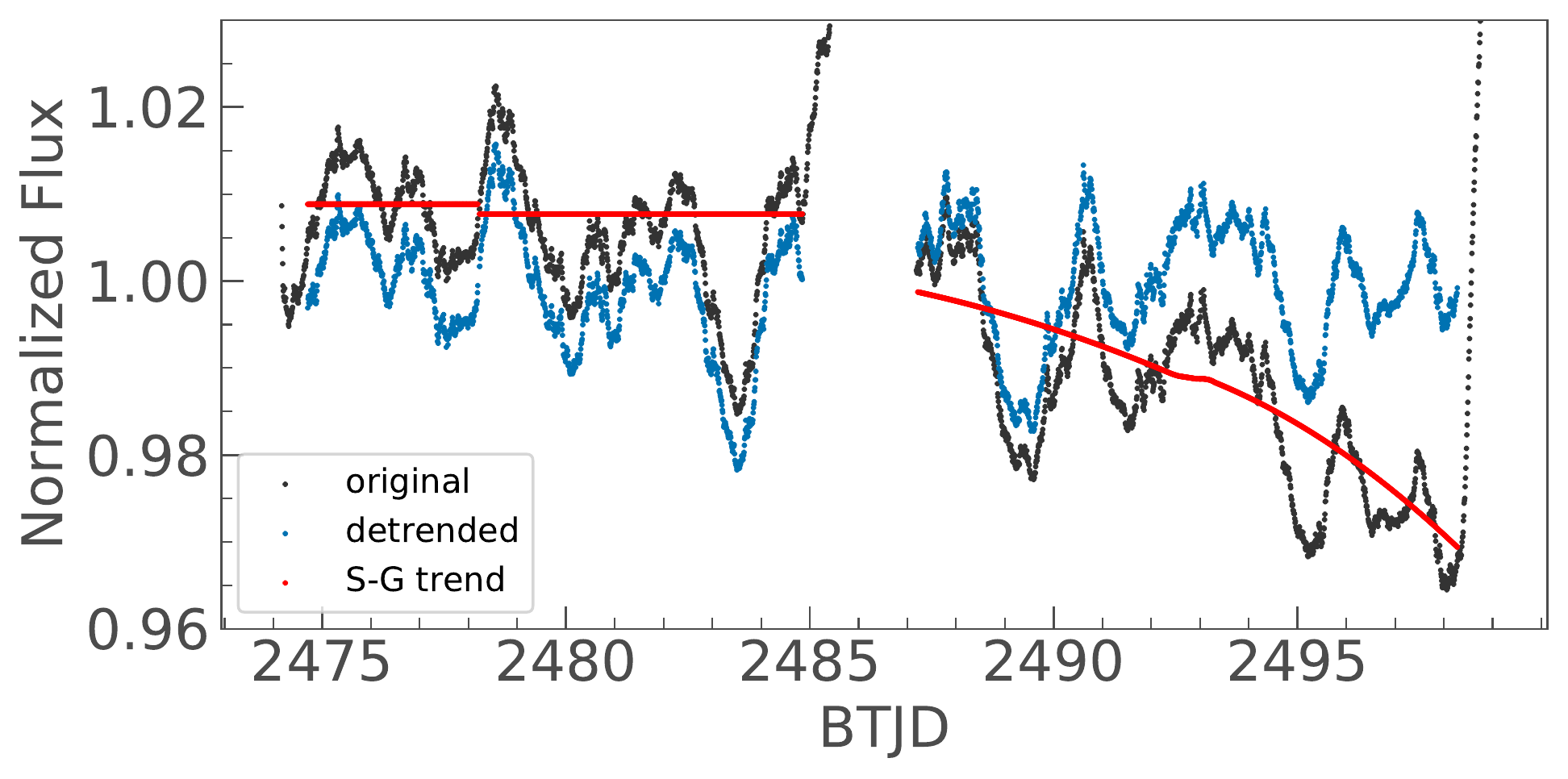}}
  \subfigure[Sector\,45]{\includegraphics[width=7.8cm, trim=2.75cm 0.1cm 0.2cm 0.1cm, clip]{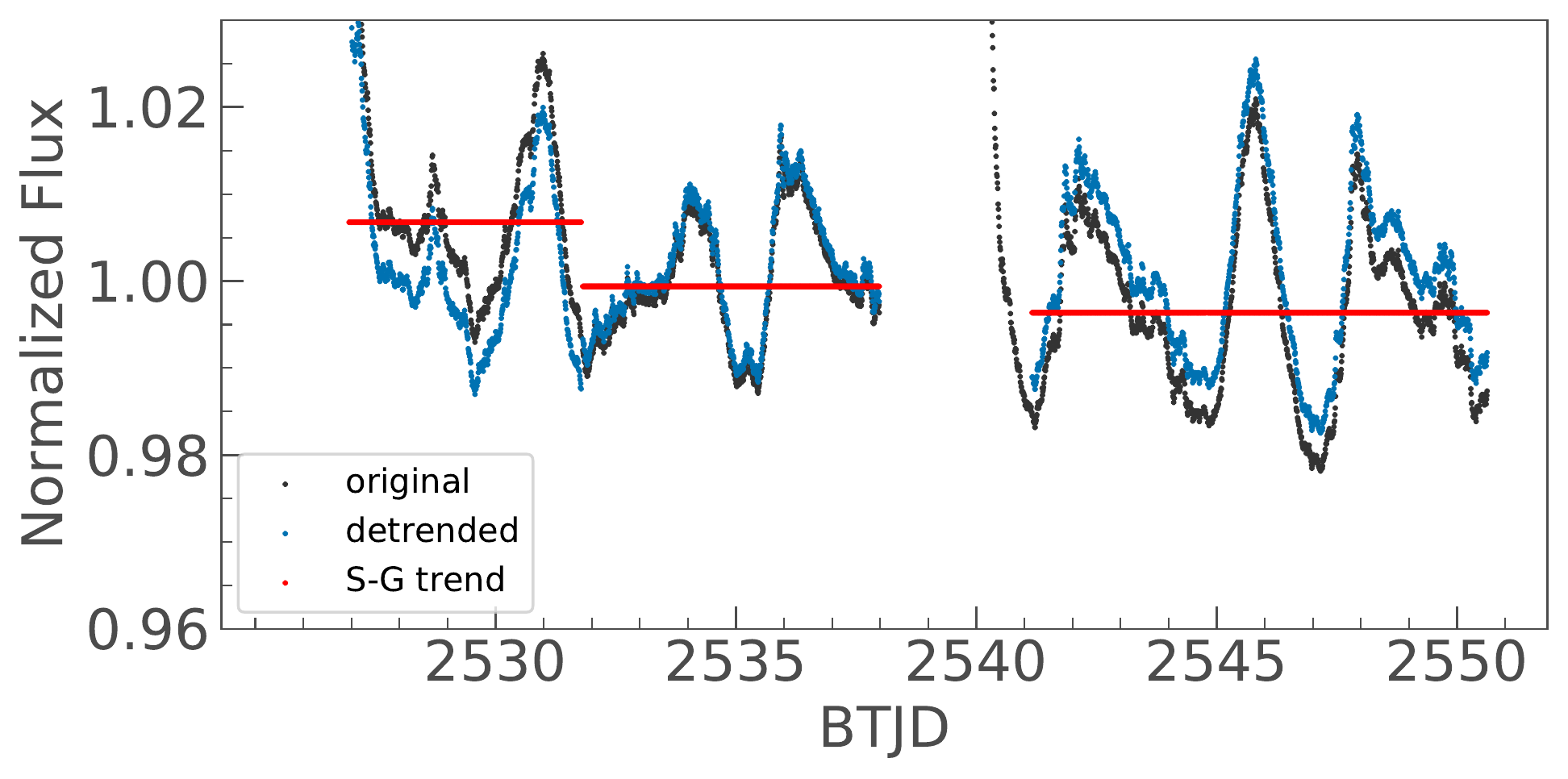}}
  \caption{Corrected and original light curves shown in blue and black, respectively. The red curves are the trend of the light curves obtained with the S-G filter. (a)-(d) represent the light curve of four different sectors respectively.}%????
  \label{lc_RC}%??????
\end{figure*}

One method is utilizing the Savitzky-Golay (hereafter S-G) filter \citep{1964AnaCh..36.1627S}, which is commonly used for detrending. A proper window length is a vital factor for the S-G filter. If the window length is too small, it will cause over-detrending of the light curves. However, if the window length is too large, the systematic trends cannot be completely removed. Our multiple tests suggest that the light curves can be well detrended when the window lengths are 501, 1851, 1501, and 1501 for the Sectors 06, 32, 43, and 45, respectively. As shown in Figure\,\ref{lc_RC}, the red curves display the trending of the light curves in the four sectors. The black and blue curves are the original and detrended light curves, respectively.

Another method is based on the linear regression technique (LR) provided by {\tt Lightkurve} \footnote{https://docs.lightkurve.org/tutorials/2-creating-light-curves/ 2-3-removing-scattered-light-using-regressioncorrector.html}. This method identifies a set of trends in the pixels surrounding the target star (i.e., the pixels outside the mask), and performs linear regression to create a combination of these trends that effectively models the systematic noise introduced by the spacecraft's motion and then subtracts the noise model from the uncorrected light curve. Too small of an image size is not conducive to detect the trend, and furthermore fails to perform effective detrending, thus, a larger number of target pixels (e.g., 50\,pixel $\times$ 50\,pixel for this study) should be used when adopting the method of linear regression. 

We use the Combined Differential Photometric Precision (CDPP) metric to compare the performance of detrending the light curves in different ways. CDPP is a measure of the precision achieved by the processing that includes statistical techniques and accommodation for systematic errors. A smaller CDPP value indicates better precision. Table\,\ref{CDPP} specifies the values of the CDPP of the detrended light curves of the four sectors from the three separate methods, i.e., the S-G filter, LR, and QLP. The values of CDPP suggest that the method of the S-G filter has the best performance in detrending the light curves.

Finally, we inspect the detrended light curves from the above two methods and from the QLP, and find that the effect of the scattered light in the beginning or end of the light curves is not completely removed in all three methods. Therefore, we remove the abnormal parts in the beginning or end of the light curves.

\begin{table}[H]
\footnotesize
\begin{threeparttable}
\caption{CDPP detrending evaluation.}\label{CDPP}
\doublerulesep 0.1pt \tabcolsep 13pt
\begin{tabular}{cccc}
\hline
\multirow{2}{*}{Sector} & \multicolumn{3}{c}{CDPP [ppm]}         \\ 
\cline{2-4} 
                        & S-G     & LR & QLP     \\ \hline
06                      & 1364.47 & 1397.76           & 2696.64 \\
32                      & 714.56  & 748.91            & 777.93  \\
43                      & 953.57  & 1083.01           & 1055.08 \\
45                      & 913.87  & 886.92          & 1120.43  \\ \hline
\end{tabular}

%\begin{tablenotes}
%\item[]{$^1$ Unit ppm is parts per million.}
%\end{tablenotes}
%\\{\bf COMMENT: ADD A FOOTNOTE EXPANDING THE ABBREVIATION PPM.}
\end{threeparttable}
\end{table}

\section{Periodic analysis}
\label{Period analysis}
\begin{figure*}[htp]
  \centering
  \subfigure[Sector\,06]{\includegraphics[width=0.48\textwidth, trim=0.3cm 0.3cm 0.3cm 0.3cm, clip]{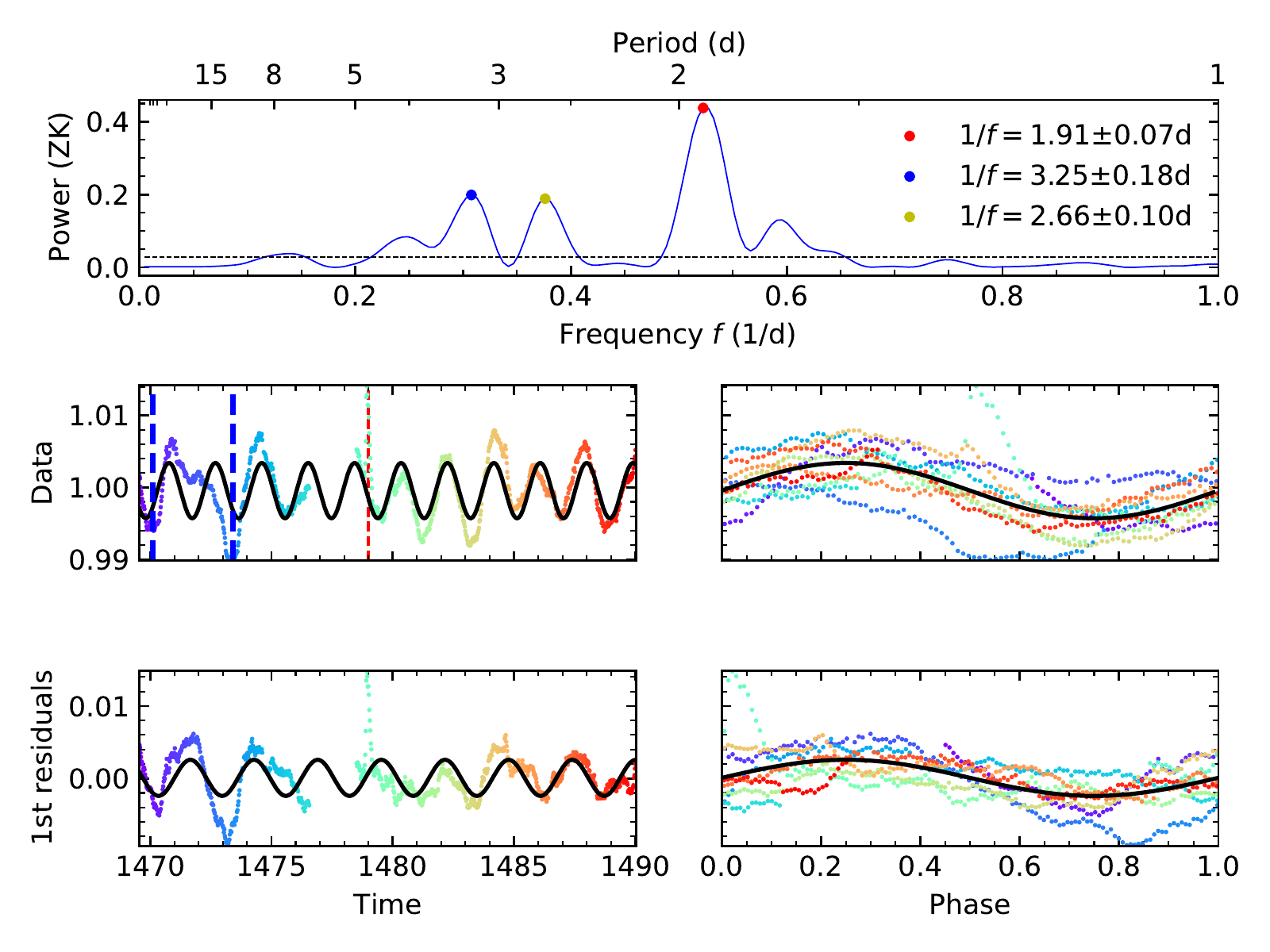}}
  \subfigure[Sector\,32]{\includegraphics[width=0.48\textwidth, trim=0.3cm 0.3cm 0.3cm 0.3cm, clip]{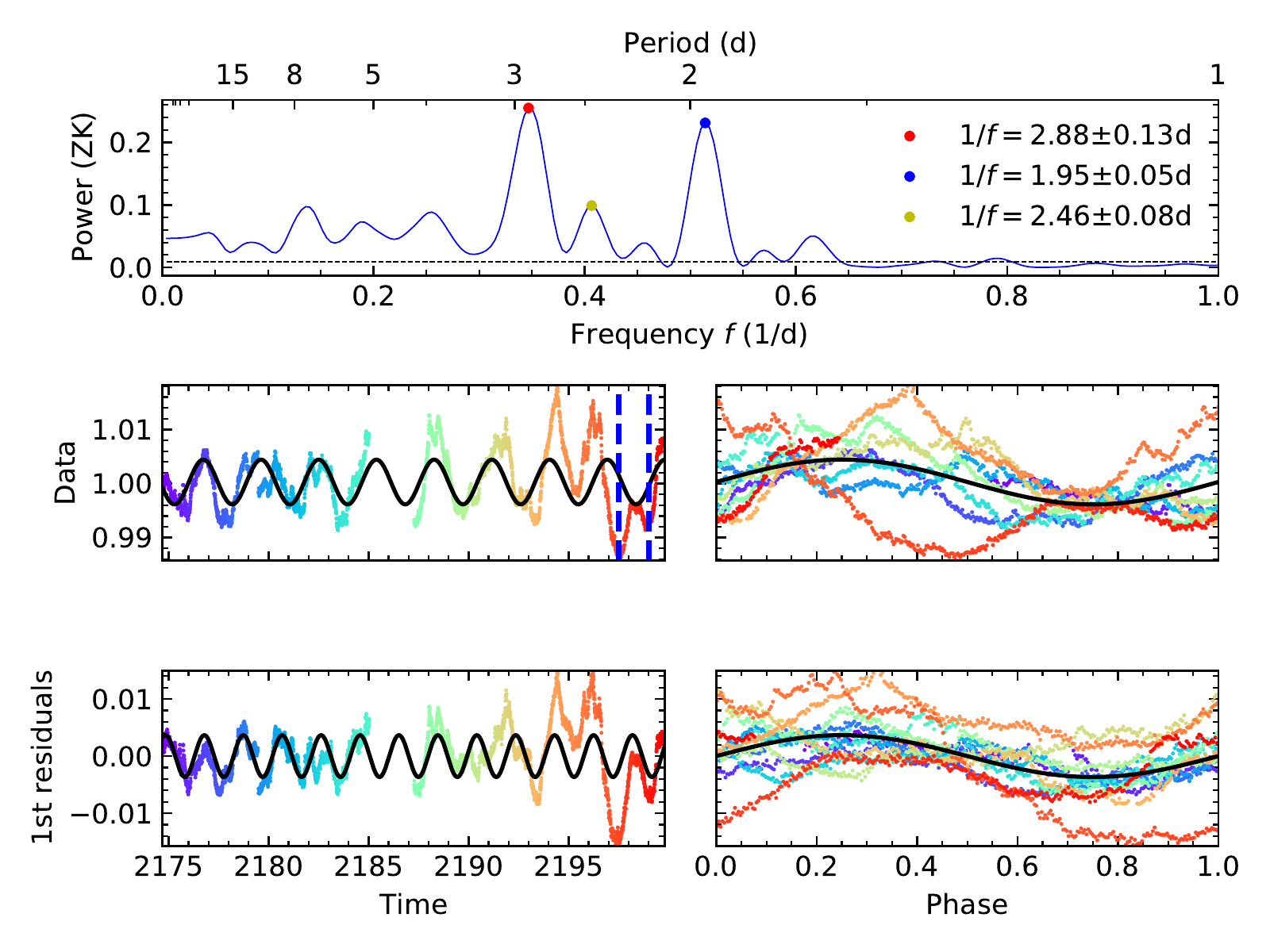}}
  \subfigure[Sector\,43]{\includegraphics[width=0.48\textwidth, trim=0.3cm 0.3cm 0.3cm 0.3cm, clip]{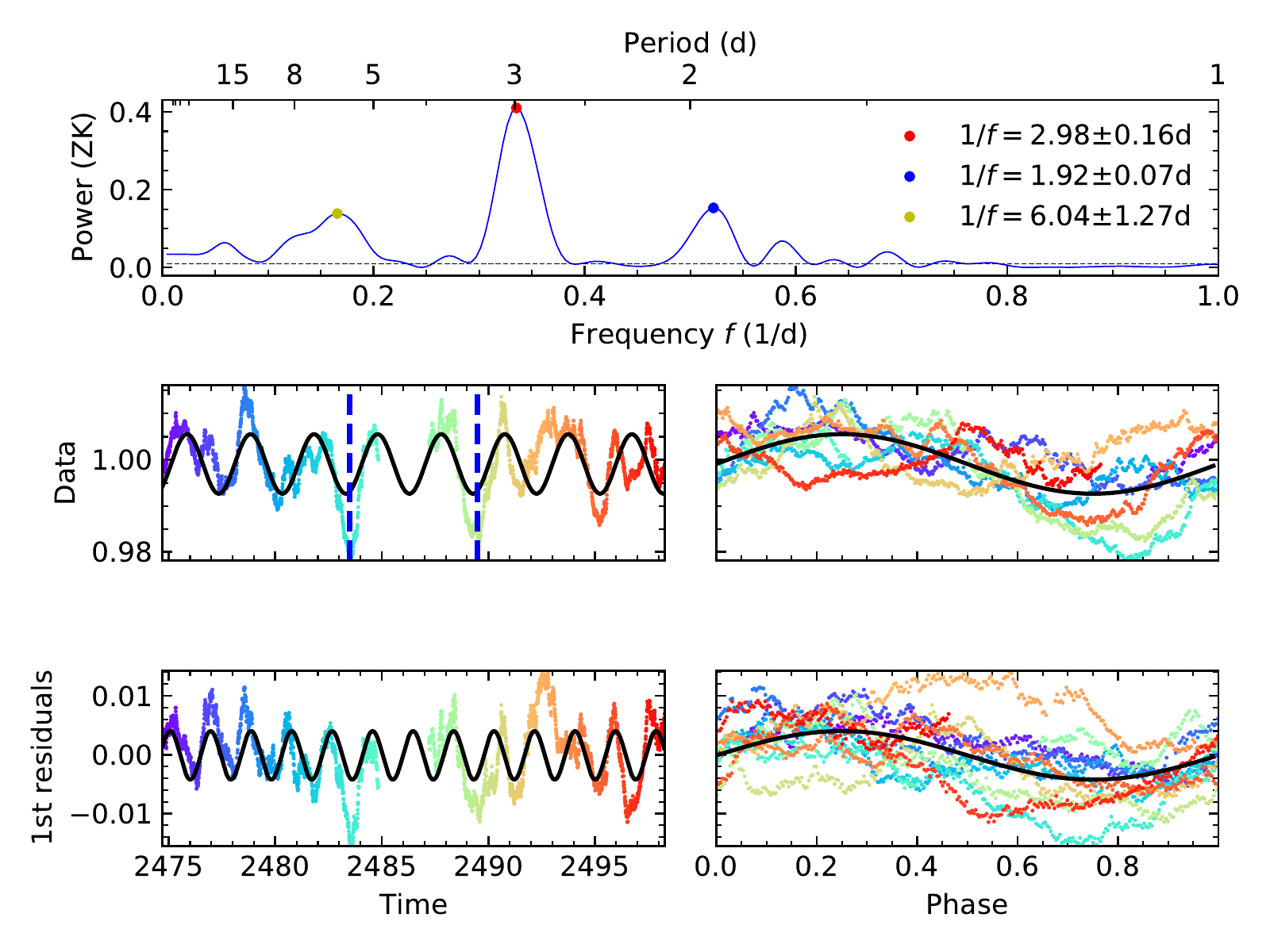}}
  \subfigure[Sector\,45]{\includegraphics[width=0.48\textwidth, trim=0.3cm 0.3cm 0.3cm 0.3cm, clip]{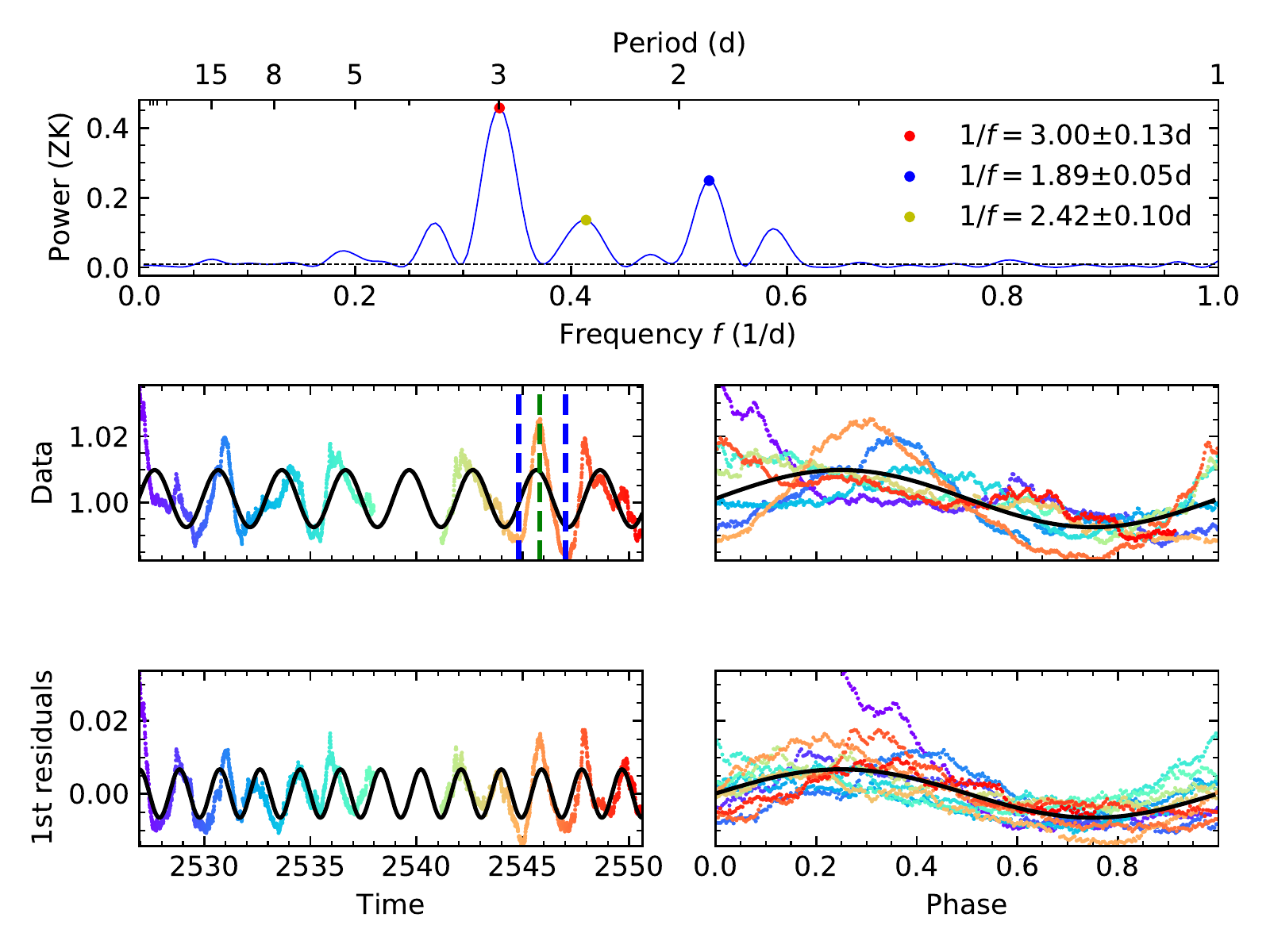}}
  \caption{GLS periodograms of GW~Ori for four sectors (panels (a)$-$(d)). Top sub-panels display the power spectrum, and the red, blue, and yellow dots mark the locations of the first three peaks of the power spectrum. The gray dotted lines are the false alarm probability of 0.1\%. The middle sub-panels illustrate the observed light curves (left), and the corresponding phase diagrams (right), both of which are overlapped by a sinusoidal wave (black) with the strongest frequency. The bottom sub-panels are similar with the middle ones, except that the input data is the first residual for each sector. The colors of the middle and bottom sub-panels represent each period corresponding to different times. ``Dipper'' events are marked with thick blue vertical dashed lines; the flare is marked with a thin red vertical dashed line in panel (a), and the ``Brightening'' event is marked by a thick green vertical dashed line in panel (d).}%????
  \label{gls}%??????
\end{figure*}

To search for reliable periodic signals from the triple system GW~Ori, we apply the GLS periodogram to the S-G detrended light curves in the four sectors. The GLS is a useful tool to study periodicity and perform frequency analysis of time series, providing accurate frequencies and spectral intensity measurements. We carry out GLS using the program of \cite{pya}, and illustrate our results for the four sectors in panels (a)$-$(d) of Figure\,\ref{gls}.

As shown in the top sub-panel for each sector, the red, blue, and yellow dots mark the locations of the first three peaks ordered by the strength of the signal in the power spectra from strongest to weakest. The corresponding period values are listed in the top-right corner. In all four sectors, the two most significant signals have average periods of $\sim$1.92$\pm0.06$\,d and $\sim$3.02$\pm0.15$\,d, and are considerably stable within 1-$\sigma$. 
In addition, we detect a third signal with an average period of 2.5\,d in Sectors 06, 32, and 45. This signal is not quite stable, since it does not occur in Sector\,43. The third peak in Sector\,43 takes place at a frequency corresponding to a period of 6.0\,d, which happens to be two times of the period of 3.0\,d. This suggests that they are very likely from the same signal, which also is probably the same signal with a period of $5.0-6.7$\,d detected by \cite{2014A&A...570A.118F}. It is worth noting that the periods of 3.25$\pm$0.18\,d and 2.66$\pm$0.10\,d detected in Sector\,06 are slightly larger than those in the other sectors. However, the differences are not statistically significant; the signals are the same within 1-$\sigma$ in all four sectors. 

The middle row left column sub-panels of Figure\,\ref{gls} illustrate the detrended light curve for each sector. The middle row right column sub-panels display the corresponding phase diagram for each sector. The colors represent each period corresponding to different times. Both middle row sub-panels overplot in black the sinusoidal wave (i.e., the primary wave) characterized by the strongest frequency detected in the power spectrum (the red dot in the top sub-panels).

The bottom two sub-panels are similar to the middle sub-panels, however, they demonstrate the first residual signal (left) and the corresponding phase (right) for each sector. The first residual signal is defined as the remaining signal after the sinusoidal wave with the strongest frequency is removed from the detrended light curve. The black curve is the sinusoidal wave (i.e., the secondary wave) with the second strongest frequency detected in the power spectrum (the blue dot in the top sub-panels). 

\begin{figure*}[htp]
  \centering
  \subfigure[Sector\,06]{\includegraphics[width=0.45\textwidth, trim=0.1cm 0.3cm 0.1cm 0.1cm, clip]{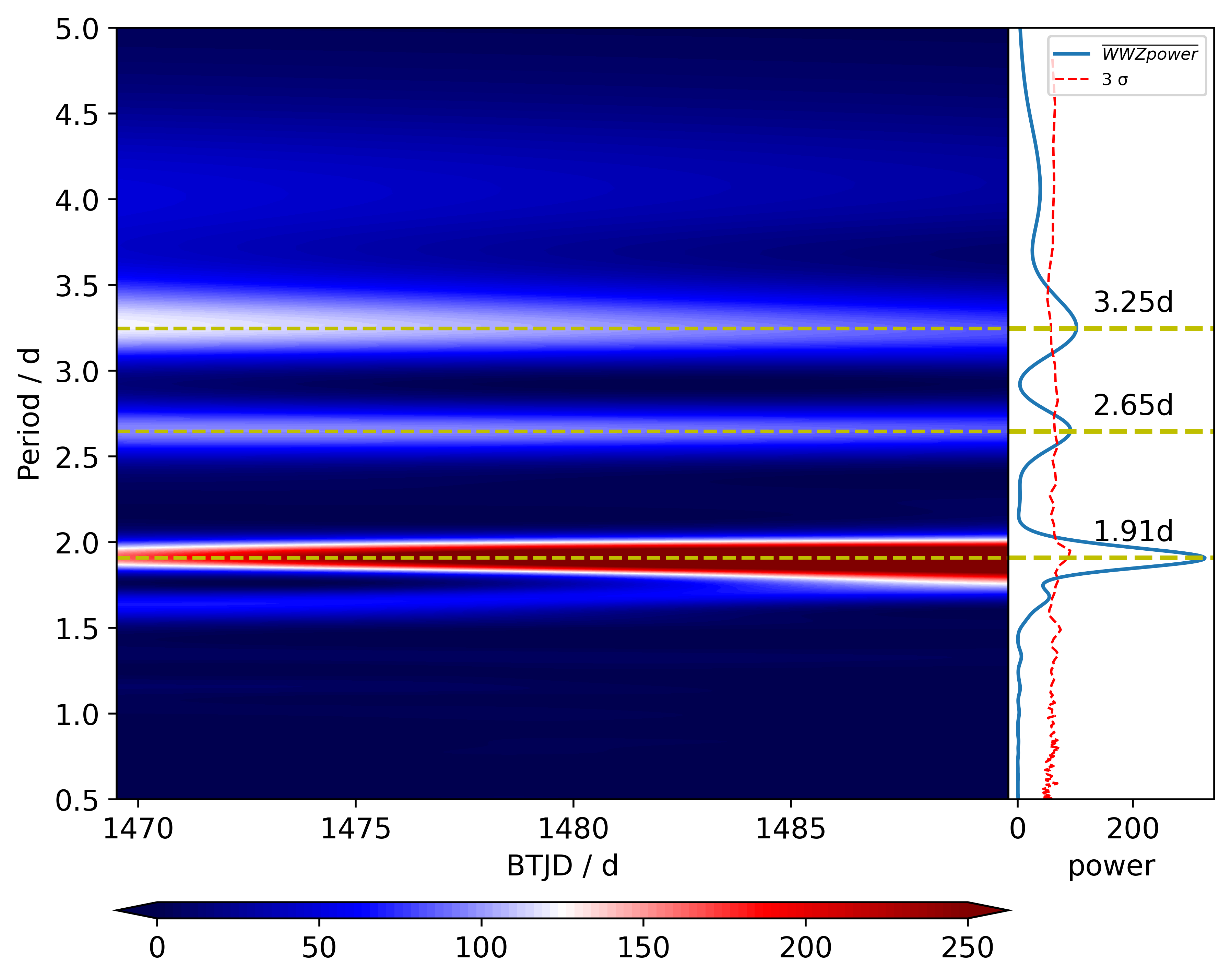}}
  \subfigure[Sector\,32]{\includegraphics[width=0.45\textwidth, trim=0.1cm 0.3cm 0.1cm 0.1cm, clip]{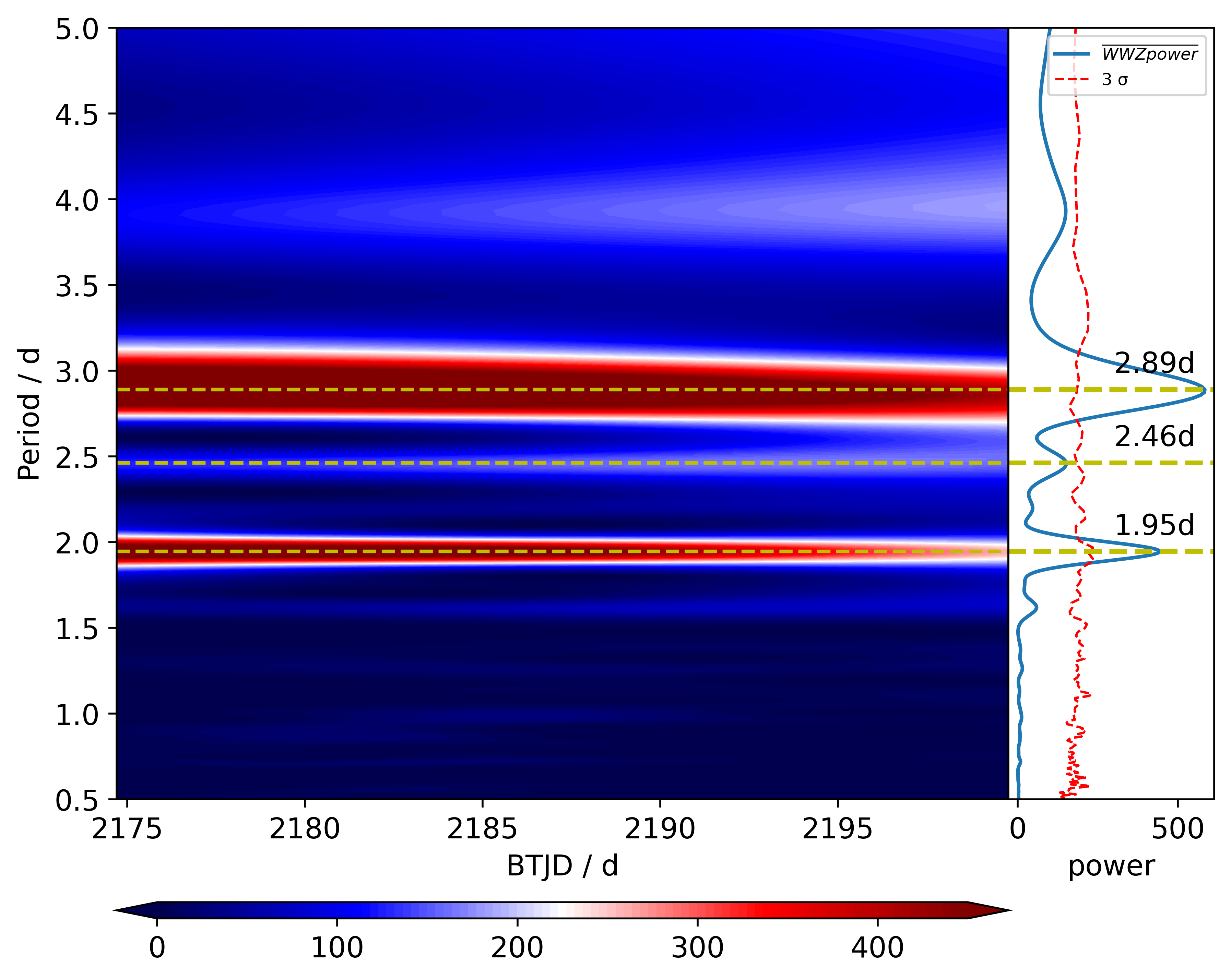}}
  \subfigure[Sector\,43]{\includegraphics[width=0.45\textwidth, trim=0.1cm 0.3cm 0.1cm 0.1cm, clip]{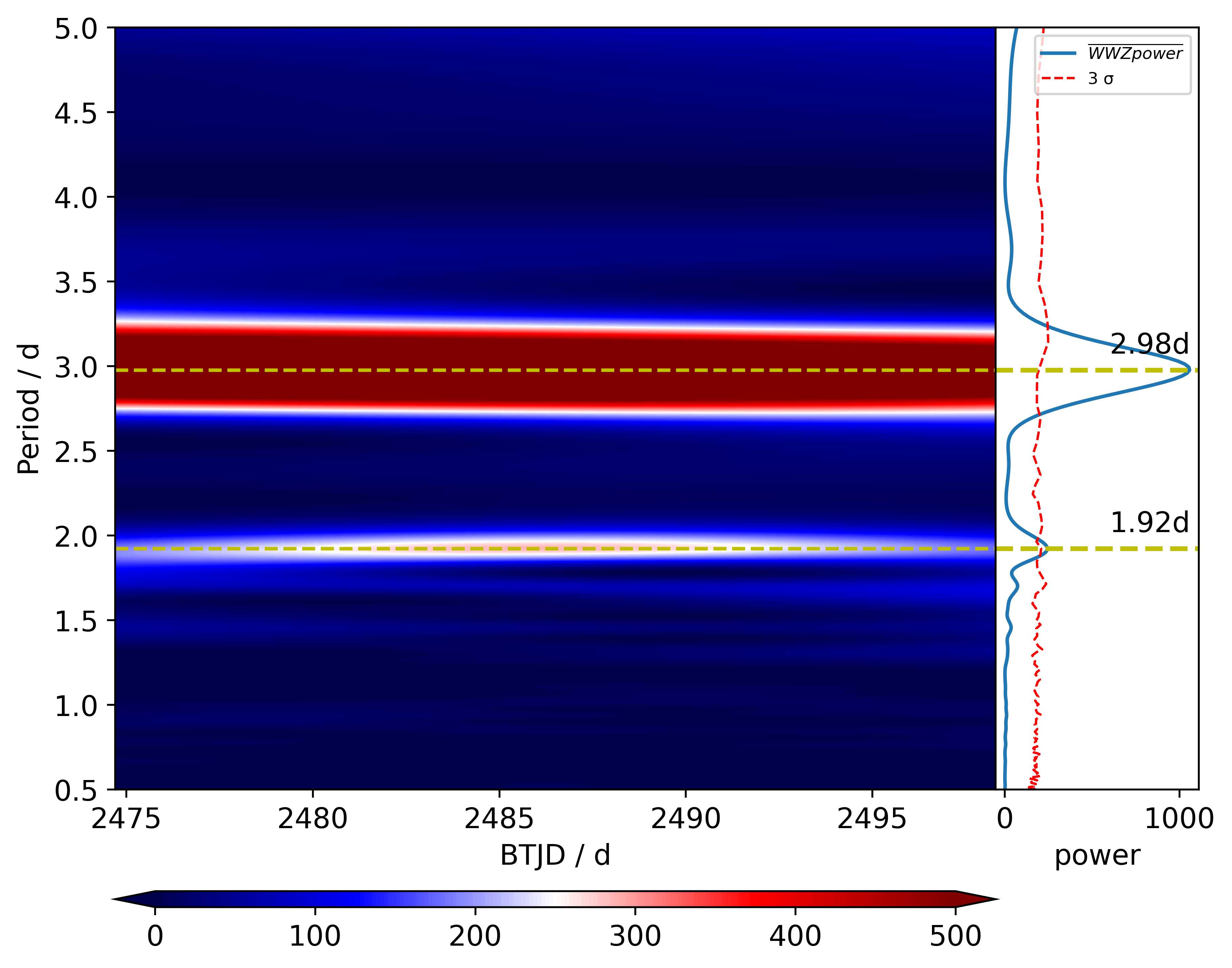}}
  \subfigure[Sector\,45]{\includegraphics[width=0.45\textwidth, trim=0.1cm 0.3cm 0.1cm 0.1cm, clip]{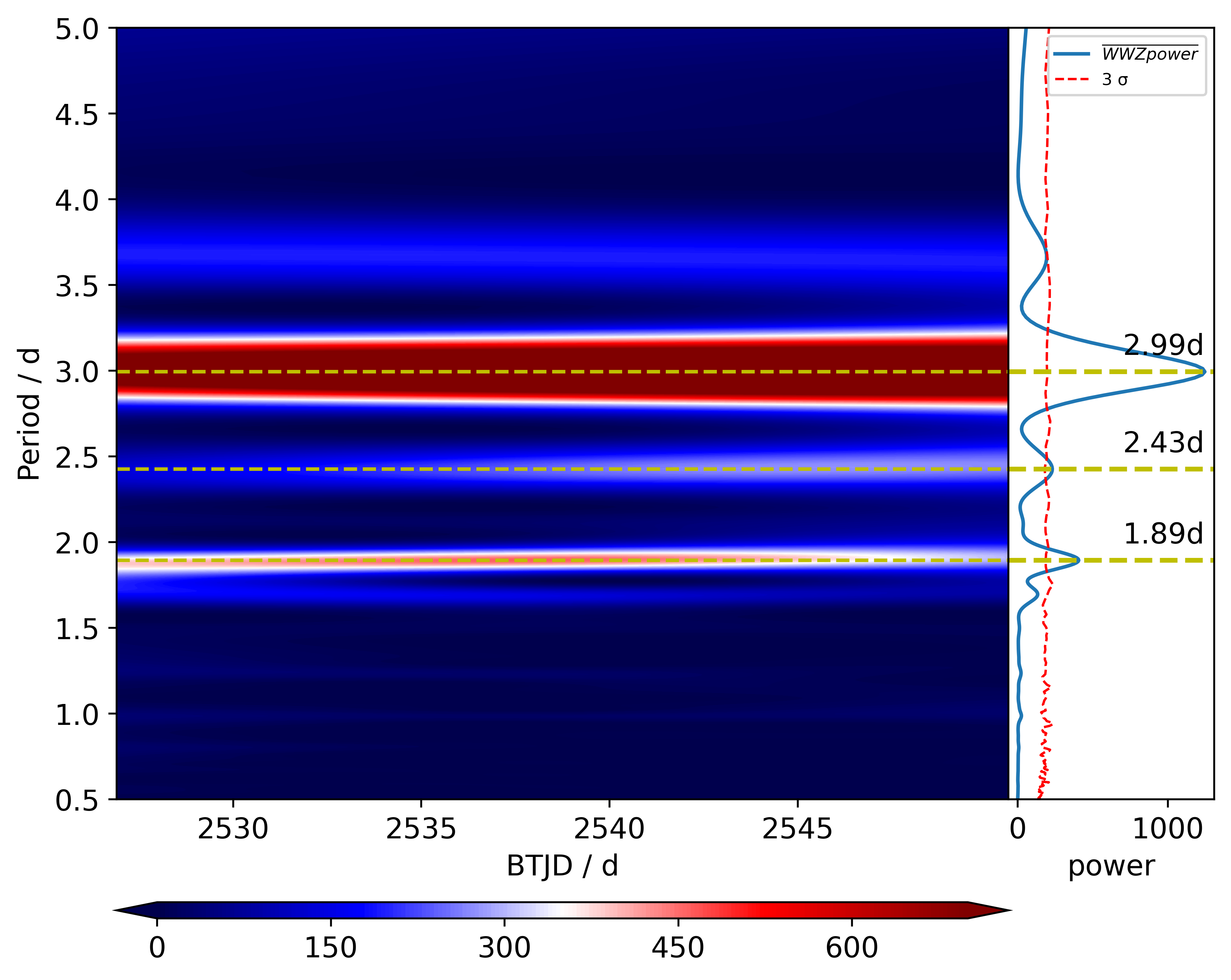}}
  \caption{Left sub-panels: WWZ power from GW~Ori for four sectors (panels (a)-(d)). Right sub-panels: average WWZ power. The most significant periodic signals (three in
Sectors\,06, 32, and 45,
and two in Sector\,43) are marked with yellow dashed lines (main and side panels). The red dashed curves (right side panels) are the 3-$\sigma$ confidence.}%????
  \label{wave}%??????
\end{figure*}

We also employ one of the most widely used methods to the S-G
detrended light curves, the Weighted Wavelet Z-transform \citep[WWZ,][]{1996AJ....112.1709F}, to obtain the power spectra of the GW~Ori light curves. 
The peak values of 1.9\,d and 3.0\,d of the two periodic signals are clearly detected in WWZ powers from all four sectors, and the signal with a period of 2.5\,d also can be seen throughout the observations in Sectors\,06, 32, and 45 (Figure\,\ref{wave}). To estimate the confidence level more robustly, we generate $10^{4}$ artificial light curves using the {\tt REDFIT} method \citep{SCHULZ2002421}. The simulation results show that the first two signals exceed a 3-$\sigma$ false-alarm level, which indicates that these two signals are stable. The third signal only slightly exceeds a 3-$\sigma$ false-alarm level (red dashed curves in the side panels of Figure\,\ref{wave}).

Moreover, we find several flux dips (1$\%$$-$2$\%$ depth) in the light curves, which are marked with the thick blue vertical dashed lines in the middle row left column sub-panels of Figure \ref{gls}. These dips are possibly related to ``Dipper'' events \citep{Cody2014}. We calculate the full width at half maximum (FWHM) as the timescales of each ``Dipper'' event using a Gaussian fit in the second residuals (i.e., the light curves of GW~Ori after deducting its two sinusoidal waves). The values of BTJD (occurrence time of the ``Dipper'') and FWHM are specified in Table\,\ref{event}. 
Benefiting from the high accuracy of TESS observations, we also find a possible ``Brightening'' event \citep[1$\%$$-$2$\%$ increase in flux,][]{Cody2014} apart from the ``Dipper'' events, which is marked with the thick green vertical dashed line at 2545.8\,d of Sector\,45 (FWHM is 0.41\,d). The brightening event shows a sudden increase in flux (within one day) followed by a decrease in flux on a similar time scale. We discuss these features in Section \ref{Aperiodic}.

We also find an obvious flare event at BTJD$\sim$1479.0\,d in the light curve of Sector\,06, which is marked with the thin red dashed line in the middle left column sub-panel of panel (a) in Figure \ref{gls}.
%======================================================

\begin{table}[H]
\footnotesize
\begin{threeparttable}
\caption{Short-duration ``Dipper'' events in BTJD.}%????
\label{event}%??
\doublerulesep 0.1pt \tabcolsep 13pt
\begin{tabular}{ccccc}
\hline
\multirow{2}{*}{Sector} & \multicolumn{4}{c}{BTJD(day)}                                   \\ \cline{2-5} 
                        & \multicolumn{2}{c|}{Dipper}          & \multicolumn{2}{c}{FWHM} \\ \hline
06                      & 1470.1 & \multicolumn{1}{c|}{1473.4} & 0.97        & 0.87       \\
32                      & 2195.4 & \multicolumn{1}{c|}{2197.5} & 1.08        & 0.61       \\
43                      & 2483.5 & \multicolumn{1}{c|}{2489.5} & 0.86        & 1.18       \\
45                      & 2544.8 & \multicolumn{1}{c|}{2547.0} & 0.88        & 0.60       \\ \hline
\end{tabular}
\begin{tablenotes}
\item[]{Note: Second and third columns are the occurrence times of the ``Dippers'' in each sector. Fourth and fifth columns are the ``Dipper'' timescales.}
\end{tablenotes}
\end{threeparttable}
\end{table}

To summarize, we detect two significant signals with average periods of 1.92 $\pm$ 0.06\,d and 3.02 $\pm$ 0.15\,d, and one possible (the third, weak and not quite stable) periodic signal with an average period of 2.51 $\pm$ 0.09\,d using two different methods (i.e., GLS and WWZ) from the light curves of GW~Ori observed by TESS. In addition, we find several aperiodic features in the light curves. We discuss the possible origins of these periodic and aperiodic signals in the following Section \ref{Discussion}.

\section{Discussion}
\label{Discussion}

In this section, we discuss the reliability and possible origins of the periodic and aperiodic signals detected in the light curves of GW Ori introduced in Section \ref{Period analysis}, and then analyze the physics behind them.

\subsection{Reliability and possible origins of the periodic signals}
\label{Possibility}
%==============================
To check the reliability of the periodic signals that we detect in Section\,\ref{Period analysis}, we try several cases to re-analyze the light curves of GW~Ori.

(1) We use small mask sizes with different threshold values, e.g., 20, 18, 3, 23 for Sectors\,06, 32, 43, and 45, respectively, in which the brighter {\it Gaia} contaminator (12.7 mag in $G$ band) is excluded. After detrending the light curves, we carry out the GLS on the S-G detrended light curves, and find the periodic signals with periods of 3.02$\pm$0.15\,d and 1.92$\pm$0.06\,d are still clearly detectable in all the four sectors, and the third signal with a period of 2.51$\pm$0.09\,d also can be seen in Sectors\,06, 32, and 45.

(2) We carry out the GLS and WWZ (Section\,\ref{Period analysis}) analysis methods on the LR and QLP detrended light curves (Section\,\ref{detrending}) of GW~Ori. Initially in Section\,\ref{Period analysis} we present the results of carrying out GLS and WWZ on the S-G detrended light curves. Finally, we obtain the four additional results of our periodic analyses, all of which successfully detect the periodic signals of GW Ori similarly as presented in Section\,\ref{Period analysis}.

We therefore conclude that the periodic signals that we obtain in Section\,\ref{Period analysis} are reliable, which is well supported by the fact that these target signals exist in all of the above cases. Compared to the tests from this section, our results presented in Section\,\ref{Period analysis} prove to be the optimal case analysis method.

%================================

%?
%???????
What are the origins of these periodic signals? Short-period several-day signals in a binary or triple system have multiple possible origins, such as eclipsing \citep[e.g.,][]{2020MNRAS.498.6034M}, starspot modulation \citep[e.g.,][]{2020ApJ...905...67P}, pulsation \citep[e.g.,][]{2011MNRAS.415.3531B, 2020ApJ...895..124Z}, and so on. For example, the eclipsing binary KIC 8301013 has an orbital period of $P\sim4.4$\,d \citep{2011AJ....141...83P}, and also has two rotational periods ($P_{min}\sim4.2$\,d, $P_{max}\sim4.3$\,d) detected from out-of-eclipse light curves \citep{2017AJ....154..250L}. \cite{2020ApJ...905...67P} ascribed such quasi-sinusoidal variations to starspot modulation.

We easily rule out the eclipsing channel caused by orbital motion since GW~Ori has long ($\sim$242\,d for A-B, and $\sim$11 year for AB-C) orbital periods. However, it is hard to judge whether the short-period signals are from starspot modulations or pulsations. According to the stellar atmospheric parameters of GW~Ori provided in the literature, e.g., the effective temperatures for the primary ($\sim$5700$\pm$200\,K) and secondary ($\sim$4900$\pm$200\,K) components from \cite{2017ApJ...851..132C}, and the surface gravity ($\log g\sim3.3$) estimated from the stellar mass and radius \citep{2014A&A...570A.118F}, the primary and secondary components are almost located outside the instability strip \citep{2002MNRAS.333..251H} on the H-R diagram. They belong to K- and G-type stars, on which cool (caused by magnetic fields) or hot (caused by accretion) spots \citep{2021MNRAS.508.3427F} are very common phenomena like on our Sun. In addition, a cool/hot spot on the stellar surface may not be entirely stable, and this instability could be reflected in the periodic signals we detect. This is different from stellar pulsation which usually gives rise to a strictly stable period. 

In this work, we actually have detected some unstable features in the periodic signals, for instance, the period of 2.51\,d does not occur in Sector\,43, and the period of 3.25\,d in Sector\,06 is slightly larger than those in the other three sectors, even though this is not statistically significant. We also note the unstable amplitudes and phases of the main and secondary waves, i.e., neither of these are exactly consistent in all four sectors. This instability may be caused by the brightening and dimming or the appearance and disappearance of spots on the stellar surface \citep[e.g.,][figure\,9]{2017ApJ...836..200G}. Therefore, we propose starspot modulation as the likely explanation of the periodic features in the light curves. We expect the unstable features are most likely caused by individual spots appearing on the component stars of GW~Ori. For example, we attribute our 3.02$\pm$0.15\,d periodic signal as due to the GW~Ori\,A primary, as similarly suggested by \cite{2017ApJ...851..132C} who find a period of 2.93\,d.

In reality, the situation perhaps is much more complex than that expected for a young triple system with a huge circumtriple disk like GW~Ori. To unambiguously reveal the channels that cause the periodic signals and unstable periods, further observations are required.

%?
%====================================

\subsection{Periodic variability of the GW~Ori light curve}

It is hard to find the optical spectroscopic signatures of GW~Ori\,C since its predicted $V$-band flux contribution is very small ($<$5\% of the total flux), even though this component is nearly a solar-mass star \citep{2017ApJ...851..132C}. Thus, we are not sure whether the third (a weak, unstable, and not statistically significant) signal that we detect is from GW~Ori\,C. Therefore, we mainly discuss the two significant periodic signals, which likely originate from the other two components (i.e., GW~Ori A and B). However, we can not distinguish which component corresponds to which period, 1.92 $\pm$ 0.06\,d or 3.02 $\pm$ 0.15\,d.

Assuming that 3.02 and 1.92\,d correspond to the rotational periods of GW~Ori\,A and B, respectively, we then estimate the inclination $i$ of the rotation axis with the following equation \citep{2018ApJ...853L..34B}:
\begin{equation}
\sin{|i|}=\frac{1.96\times 10^{-2}\times P_{\rm gls}({\rm days}) \times v\,\sin{i({\rm km}\,{\rm s}^{-1} ) }}{R(R_{\odot })}
\label{i}
\end{equation}
%\noindent 
where $P_{\rm gls}$ is the period calculated by GLS. For GW~Ori\,A and B, the values of $v\,{\rm sin}\,i$ are 43 $\pm$ 8\,${\rm km}\,{\rm s}^{-1}$ and 50 $\pm$ 8\,${\rm km}\,{\rm s}^{-1}$, respectively (with 8\,${\rm km}\,{\rm s}^{-1}$ as the spectral resolution, \citealt{2018ApJ...852...38P}), and the values of radius $R$ are 5.90 $\pm$ 0.18\,$R_{\odot }$ and 5.01 $\pm$ 0.22\,$R_{\odot }$ \citep{2017ApJ...851..132C}, respectively. 

Using Eq.\,\ref{i}, we obtain $|i_{A}|$=25.6 $\pm$ 4.8\degree{} and $|i_{B}|$=22.1 $\pm$ 3.6\degree{}. 
 Our results indicate that the rotation axes of GW~Ori\,A and B are not aligned with the disk spin axis (i.e., $\sim$35\degree{}, \citealt{2017A&A...603A.132F}; \citealt{2017ApJ...851..132C}; \citealt{2020ApJ...895L..18B}; \citealt{2020Sci...369.1233K}), but instead are almost aligned (modulo the absolute orientation) with the orbit of A-B (i.e., $\sim$156\degree{}, \citealt{2017ApJ...851..132C}; \citealt{2020Sci...369.1233K}).

If, instead, we make the contrary assumption, i.e., GW~Ori\,A has a rotational period of 1.92\,d and GW~Ori\,B of 3.02\,d, using Eq.\,\ref{i} we then obtain $|i_{A}|$=15.9 $\pm$ 3.0\degree{} and $|i_{B}|$=36.2 $\pm$ 5.9\degree{}. We prefer the former assumption, i.e., 3.02 and 1.92\,d correspond to the rotation period of GW~Ori\,A and B, respectively, because in this case their rotation axes are almost aligned with the orbital axis of A-B, thus making the system much simpler than that of the contrary case.

\subsection{Aperiodic variability of the GW~Ori light curve}
\label{Aperiodic}

The ``Dipper'' event occurs frequently in young stars and typically appears as a significant drop in flux over a short time in the light curve \citep{Cody2014}. One possible origin is that a ``Dipper'' event results from an accretion flow raising material from the disk \citep{2017MNRAS.470..202B}. Alternatively, the event may be produced by clumpy accretion that occurs on the secondary star; in this case one of the components is eclipsed by the accretion stream during the rotational period of the binary, e.g., GW~Ori\,A and B \citep{1992PASP..104..479G,2018ApJ...852...38P}.

We compare the FWHM of possible ``Dipper'' events in each sector (Table \ref{event}) to the results of \cite{2015AJ....149..130S}. The results show that the mean FWHM of GW~Ori ``Dippers'' is about one day and is about $2-3$ times larger than those in \cite{2015AJ....149..130S}.

We check the eclipse events mentioned by \cite{2017ApJ...851..132C}, but the duration of these eclipse events (larger than 10\,d) is different from the ``Dippers'' (about one day) in the light curves of TESS. Additionally, the time interval of the ``Dipper'' does not correspond to the obvious orbital period of GW~Ori\,A-B, which may be related to the observation duration of TESS. At least in the current data, we do not find any correlations between "Dipper" events and the orbital period.

Although the ``Dipper'' events have no significant periodicity in the light curves of GW~Ori, we find that the duration of the two ``Dipper'' events in the same sector may be related to the rotational period of GW~Ori\,A or B. According to Table \ref{event}, the interval duration of 3.3\,d in Sector\,06 approaches the rotational period of GW~Ori\,A. Meanwhile, we find that the times of duration in Sectors\,32 and 45 (2.1\,d and 2.2\,d, respectively) are almost identical to the rotation period of GW~Ori\,B. We also find a time duration of 6\,d in Sector\,43. This could be a multiple of 2\,d or 3\,d, and, thus, we cannot determine whether it is related to the rotation of GW~Ori\,A or B. These intervals indicate that the aperiodic dimming is possibly caused by the obscuration of unstable accretion streams onto GW~Ori\,A or B, demonstrating that both GW~Ori\,A and B show evidence of accretion activities. 

Accretion instability is one possible channel by which a ``Brightening'' event can occur \citep{Cody2014}. The ``Brightening'' feature detected in this study is possibly a random accretion event as this feature seems to appear randomly in the light curve of GW~Ori.

We add a note of caution that the process of detrending changes the amplitude of the light curves, so it is possible that both the ``Dipper'' and ``Brightening'' features are damaged in this step.

\section{Conclusions}
\label{Conclusions}
The discovery of triple-star systems is limited and infrequent, particularly pre-main sequence triples within a nearby proximity (such as GW Ori) are even rarer.

The unique structure of GW Ori and its close proximity opens the door wide to probe numerous scientific questions. GW Ori is a hierarchical triple-star structure surrounded by three circumstellar rings (circumtriple disk). Such a unique system makes GW Ori an ideal platform to probe key topics such as the evolution of circumstellar discs (e.g., \citealt{2023MNRAS.520.6159C}), accretion rates (e.g., \citealt{2022MNRAS.514..906C}), circumtriple rings and planets (e.g., \citealt{2021MNRAS.508..392S}), dynamical interactions between triple-star systems and their circumtriple disks (e.g., \citealt{2020ApJ...895L..18B}), formation scenarios of wide binaries (e.g., \citealp{2012Natur.492..221R}; \citealp{2019MNRAS.489.5822E}), and so on.

The period is one of the most fundamental observational parameters for a binary or triple system. In particular, short-term periods (e.g., rotational modulation period) provide vital information about the geometry of the system, rotational speed, magnetic fields and accretion on stars \citep{2021MNRAS.508.3427F}. For example, measuring the rotation period of the Alpha Cen triple star system (the nearest neighbor to the Sun) has proven extremely challenging. One possible method is through accurate measurements of the strongest emission features, such as Mg\,II h and k (2800 \AA) , O\,I (1300 \AA), C\,II (1335 \AA), C\,IV (1550 \AA) and Fe\,II (2605 \AA) \citep{1995iue..prop.4969G}. Nevertheless, the rotation periods of the stellar components are vital for understanding the magnetic behavior (e.g., magnetic braking and stellar dynamo) on stars \citep{1997AAS...18912004J}, constraining stellar evolutionary models that include rotation, and studying theories of rotational evolution and redistribution of angular momentum in the interiors of solar-type stars \citep{1995iue..prop.4969G}. 

Unfortunately, due to the lack of quality light curves, there has been a long debate over the short-term periodic analysis of GW Ori. TESS has provided a nearly continuous series of FFI for more than ninety days (divided into four sectors) with high time resolution ($10-30$ minutes) and precise ($\sim$5\,mmag) photometry for GW~Ori.

We extract the original light curves for all four TESS sectors from the FFI, and perform full data preprocessing, including detrending, target mask selection, and cleaning. Applying the GLS periodogram and WWZ to the final light curves, we obtain the main results which are summarized as follows:
\begin{enumerate}
\item[(1)] We detect two significant modulation signals with periods of 3.02 $\pm$ 0.15\,d and 1.92 $\pm$ 0.06\,d, and a possible third with a period of 2.51 $\pm$ 0.09\,d from the light curves of GW~Ori. The two periodic signals are likely due to their rotational modulation caused by cool/hot (or mixed) starspots on GW~Ori A and B. 
\item[(2)] The rotational period is a bridge to elucidating the geometry of GW Ori. Given the values from the literature for the rotational velocity ($v\,{\rm sin}\,i$) and stellar radius ($R$) of components A and B, we use their periods that we measure to solve for their rotational inclination angles ($i_{A}$=25.6 $\pm$ 4.8\degree{} for GW~Ori\,A and $i_{B}$=22.1 $\pm$ 3.6\degree{} for GW~Ori\,B). These suggest that their rotational axes are almost aligned with the orbital axis of A-B. This provides new observational evidence that is key to uncovering the complex geometry of GW~Ori.

\item[(3)]We find several aperiodic features (e.g., "Dipper" and ``Brightening'' events) in the light curves. Both of the ``Dipper'' and ``Brightening'' events are likely related to unstable accretion from the disk. There are some signs that the "Dipper" features are related to rotational modulations, but the ``Brightening'' events seem to randomly occur in the light curves.

\end{enumerate}

To completely uncover the complex structure and exploit the scientific treasure of this unique triple system, more observations and future detailed investigations are required. Currently, although TESS provides supreme short-term light curves of GW~Ori, these can only be used to study the variability from short-period modulation or occultation events from rotational modulations, accretion flows or the very inner accretion disks. In the future, the long-term monitoring from TESS in combination with other surveys, e.g., Atacama Large Millimeter/submillimeter Array (ALMA) that has been the survey of choice for previous studies of GW~Ori, have great potential to further uncover the secrets hidden in this system. 

%==============================================================
{\bf Acknowledgements.}
We thank Feng Wang, Jiao Li, Yang Pan, Jian-Ning Fu, Xiao-Bin Zhang, and Sheng-Hong Gu for helpful discussions and acknowledge the National Natural Science Foundation of China (NSFC) under grant nos. 11873034, U2031202, and 12203029, the Department of Science and Technology of Hubei Province for the Outstanding Youth Fund (2019CFA087), the Cultivation Project for {\tt LAMOST} Scientific Payoff and Research Achievement of CAMS-CAS, and the science research grants from the China Manned Space Project with nos. CMS-CSST-2021-A08 and CSST Milky Way and Nearby Galaxies Survey on Dust and Extinction Project CMS-CSST-2021-A09. Funding for the TESS mission is provided by NASA's Science Mission directorate. The data supporting this article will be shared upon reasonable request sent to the corresponding author.
%==============================================================

{\bf InterestConflict.} The authors declare that they have no conflict of interest.

%\usepackage[style=numeric, sorting=ynt] {biblatex}
%\addbibresource{./bibfile.bib}
\bibliographystyle{aasjournal}
\bibliography{bibfile}

%\bibliographystyle{hunsrt}
%\bibliography{refs} % if your bibtex file is called refs.bib

\end{multicols}

\end{document}